\documentclass[12pt]{article}
\usepackage{graphicx,amsmath,amssymb,bm,subfigure}
\begin{document}

\title{Scattering of a scalar relativistic particle by the hyperbolic tangent potential}
\author{Clara Rojas}

\maketitle

\noindent{Hypergeometric functions, Klein-Gordon equation, Scattering theory}

\begin{abstract}
We solve the Klein-Gordon equation in the presence of the hyperbolic tangent potential. The scattering solutions are derived in terms of hypergeometric functions. The reflection $R$ and transmission $T$ coefficients are calculated in terms of Gamma function and, superradiance is discussed, when the reflection coefficient $R$ is greater than one.
\end{abstract}

\section{Introduction}
\label{Introduction}

The study of the solution of the Klein-Gordon equation with different potentials has been extensively studied in recent years, for bound states and scattering solutions \cite{rojas:2005,rojas:2006a,rojas:2006b,rojas:2007,arda:2011,alpdogan:2013,guo:2009a}. The scattering solutions to the  Dirac equation also have been studied for several potentials, a particular case is the Hulth\'en potential \cite{guo:2009b}.
The superradiance phenomenon, when the reflection coefficient $R$ is greater than one,  has been widely discussed. Manogue \cite{manogue:1988} discussed the superradiance on a potential barrier for Dirac and Klein-Gordon equations. Sauter \cite{sauter:1931b} and Cheng \cite{cheng:2009} have studied the same phenomenon for the hyperbolic tangent potential with the Dirac equation. 
Superradiance for the Klein-Gordon equation with this particular potential has been studied for Cheng too \cite{cheng:2009}. The hyperbolic tangent potential even wake interest, recently has discussed the bound states of scalar particle in presence of the truncated hyperbolic tangent potential \cite{gonzalez:2005} and the hyperbolic tangent potential \cite{tian:2009}.

In this paper we have calculated the scattering solutions of the Klein-Gordon equation in terms of hypergeometric functions in presence of the hyperbolic tangent potential. The reflection $R$ and transmission $T$ coefficients are calculated in terms of Gamma function. The behaviour of the reflection $R$ and transmission $T$ coefficients is studied for five different regions of energy. We have observed for some region that  $R>1$ and $T<0$, so the phenomenon of superradiance is observed in this potential \cite{cheng:2009, wagner:2010}.

This paper is organized of the following way. Section \ref{Klein-Gordon} shows the one-dimensional Klein-Gordon equation. In section \ref{tanh} the hyperbolic tangent potential is shown. Section \ref{scattering} shows the scattering solutions and the behaviour of the reflection $R$ and transmission $T$ coefficients. Finally, in section \ref{conclusion} conclusions are discussed.

\section{The Klein-Gordon equation}
\label{Klein-Gordon}

 The one-dimensional Klein-Gordon equation to solve is, in natural units $\hbar=c=1$ \cite{greiner:1987}

\begin{center} 
\begin{equation}
\label{klein}
\frac{d^2\phi(x)}{dx^2}+\left\{\left[E-V(x) \right]^2-m^2 \right\}\phi(x)=0, \tag*{(1)}
\end{equation}
\end{center}
where $E$ is the energy, $V(x)$ is the potential and, $m$ is the mass of the particle.

\section{The hyperbolic tangent potential}
\label{tanh}

\begin{figure}[htbp]
\begin{center}
\includegraphics[scale=0.50]{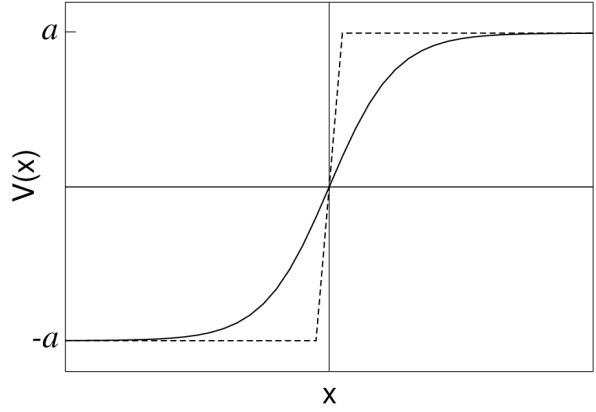}
\caption{\label{fig_pot}{Hyperbolic tangent potential for $a=5$ with $b=2$ (solid line) and $b=50$ (dotted line).}}
\end{center}
\end{figure}

The hyperbolic tangent potential is defined as 

\begin{equation}
\label{potential}
V(x)=a\,\tanh(b\,x), \tag*{(2)}
\end{equation}
where $a$ represents the height of the potential and $b$ gives the smoothness of the curve. The form of the hyperbolic tangent potential is showed in the Fig. \ref{fig_pot}. From Fig. \ref{fig_pot} we can note that the hyperbolic tangent potential reduces to a step potential for $b \rightarrow \infty$.

\section{Scattering States}
\label{scattering}

In order to consider the scattering solutions, we solve the differential equation

\begin{equation}
\label{eq_x}
\frac{d^2\phi(x)}{dx^2}+\left \{ \left[E-a\tanh(bx)\right]^2-m^2 \right \}\phi(x)=0. \tag*{(3)}
\end{equation}

\medskip
On making the substitution $y=-e^{2bx}$ , Eq. \ref{eq_x} becomes

\begin{equation}
\label{eq_y)}
4b^2 y \frac{d}{dy}\left[y\frac{d\phi(y)}{dy}\right]+\left[\left(E+a\frac{1+y}{1-y} \right)^2-m^2 \right]\phi(y)=0. \tag*{(4)}
\end{equation}

\medskip
Putting $\phi_(y)=y^\alpha(1-y)^\beta f(y)$, Eq. \ref{eq_y)} reduces to the hypergeometric differential equation
 
\begin{equation}
\label{eq_hyper}
y(1-y)f''+[(1+2\alpha)-(2\alpha+2\beta+1)y]f'-(\alpha+\beta-\gamma)(\alpha+\beta+\gamma)f=0, \tag*{(5)}
\end{equation}

where the primes denote derivate with respect to $y$ and the parameters $\alpha$, $\beta$, and $\gamma$ are

\begin{align}
\label{alpha}
\alpha &= i\nu \,\,\, \textnormal{with} \,\,\, \nu=\frac{\sqrt{(E+a)^2-m^2}}{2b},  \tag*{(6)}\\
\label{beta}
\beta &= \lambda \,\,\, \textnormal{with} \,\,\, \lambda=\frac{b+\sqrt{b^2-4a^2}}{2b},  \tag*{(7)}\\
\label{gamma}
\gamma &= i\mu \,\,\, \textnormal{with} \,\,\, \mu=\frac{\sqrt{(E-a)^2-m^2}}{2b}. \tag*{(8)}
\end{align}

\medskip
Eq. \ref{eq_hyper} has the general solution in terms of Gauss hypergeometric functions $_2F_1(\mu,\nu,\lambda;y)$ \cite{abramowitz:1965}

\begin{align}
\label{sol_y}
\nonumber
 \phi(y)&=C_1 y^\alpha \left(1-y\right)^\beta\, _2F_1\left(\alpha+\beta-\gamma,\alpha+\beta+\gamma,1+2\alpha;y\right)\\
 &+C_2 y^{-\alpha} \left(1-y\right)^\beta\, _2F_1\left(-\alpha+\beta-\gamma,-\alpha+\beta+\gamma,1-2\alpha;y\right) \tag*{(9)}.
\end{align}

\medskip
In terms of variable $x$ Eq. \ref{sol_y} becomes

\begin{align}
\label{sol_x}
\nonumber
\phi(x)&=c_1 \left(-e^{2bx}\right)^{i\nu} \left(1+e^{2bx}\right)^\lambda\, _2F_1\left(i\nu+\lambda-i\mu,i\nu+\lambda+i\mu,1+2 i\nu;-e^{2bx}\right)\\
\nonumber
 &+c_2 \left(-e^{2bx}\right)^{-i\nu} \left(1+e^{2bx}\right)^\lambda\, _2F_1\left(-i\nu+\lambda+i\mu,-i\nu+\lambda-i\mu,1-2 i\nu;-e^{2bx}\right). \tag*{(10)}
\end{align}

\medskip
From Eq. \ref{sol_x} the incident and reflected waves are

\begin{equation}
\label{phi_inc}
\phi_{\textnormal{inc}}(y)=d_1 \left(1+e^{2bx}\right)^\lambda e^{2ib\nu x}\, _2F_1\left(i\nu+\lambda-i\mu,i\nu+\lambda+i\mu,1+2 i\nu;-e^{2bx}\right) \tag*{(11)}.
\end{equation}

\begin{equation}
\label{phi_ref}
\phi_{\textnormal{ref}}(y)=d_2 \left(1+e^{2bx}\right)^\lambda e^{-2ib\nu x}\, _2F_1\left(-i\nu+\lambda+i\mu,-i\nu+\lambda-i\mu,1-2 i\nu;-e^{2bx}\right) \tag*{(12)}.
\end{equation}

\medskip
Using the relation \cite{abramowitz:1965}

\begin{align}
 \label{relation}
 \nonumber
 _2F_1(a,b,c;z)&=\frac{\Gamma(c)\Gamma(b-a)}{\Gamma(b)\Gamma(c-a)}(-z)^{(-a)}\,_2F_1(a,1-c+a,1-b+a;z^{-1})\\
 \nonumber
 &+\frac{\Gamma(c)\Gamma(a-b)}{\Gamma(a)\Gamma(c-b)}(-z)^{(-b)}\,_2F_1(b,1-c+b,1-a+b;z^{-1}). \tag*{(13)}
\end{align}

\medskip
The transmited wave becomes

\begin{equation}
\label{phi_trans}
\phi_{\textnormal{trans}}(x)= d_3 e^{-2b\lambda x} \left(1+e^{2bx}\right)^\lambda e^{2ib\mu x}\, _2F_1\left(i\nu+\lambda-i\mu,-i\nu+\lambda-i\mu,1-2 i\mu;-e^{-2bx}\right). \tag*{(14)}
\end{equation}

\medskip
As the incident  wave is equal to the sum of the  transmitted wave and the reflected wave

\begin{equation}
 \label{sumatoria}
 \phi_{\textnormal{inc}}(x)=A\,\phi_{\textnormal{trans}}(x)+B\,\phi_{\textnormal{ref}}(x). \tag*{(15)}
\end{equation}

\medskip
We used again the relation \ref{relation} and the equation for $\phi_{\textnormal{trans}}(x)$ to find

\begin{equation}
\label{phi_inc_def}
\phi_{\textnormal{inc}}(x)= A \left(1+e^{2bx}\right)^\lambda e^{2ib\nu x}\, _2F_1\left(i\nu+\lambda-i\mu,i\nu+\lambda+i\mu,1+2 i\nu;-e^{2bx}\right).\tag*{(16)}
\end{equation}

\begin{equation}
\label{phi_ref_def}
\phi_{\textnormal{ref}}(x)=B \left(1+e^{2bx}\right)^\lambda e^{-2ib\nu x}\, _2F_1\left(-i\nu+\lambda+i\mu,-i\nu+\lambda-i\mu,1-2 i\nu;-e^{2bx}\right). \tag*{(17)}
\end{equation}

\medskip
Where the coefficients $A$ and $B$ in Eqs. \ref{phi_inc_def} and \ref{phi_ref_def} can be expressed in terms of the Gamma function as

\begin{equation}
 \label{A}
 A=\frac{\Gamma(1-2i\mu)\Gamma(-2i\nu)}{\Gamma(-i\nu+\lambda-i\mu)\Gamma(1-i\nu-\lambda-i\mu)}. \tag*{(18)}
\end{equation}

\begin{equation}
 \label{B}
 B=\frac{\Gamma(1-2i\mu)\Gamma(2i\nu)}{\Gamma(i\nu-\lambda-i\mu)\Gamma(1+i\nu-\lambda-i\mu)}. \tag*{(19)}
\end{equation}

\medskip
When $x\rightarrow \pm \infty$ the $V \rightarrow \pm a$ and the asymptotic behaviour of Eqs. \ref{phi_trans}, \ref{phi_inc_def} and,  \ref{phi_ref_def}    are plane waves with the relation of dispersion $\nu$ and $\mu$,

\begin{align}
\label{phi_asym}
\phi_{\textnormal{inc}}(x)&= A e^{2ib\nu x},\tag*{(20)}\\
\phi_{\textnormal{ref}}(x)&= B e^{-2ib\nu x},\tag*{(21)}\\
\phi_{\textnormal{trans}}(x)&= e^{2ib\mu x}.\tag*{(22)}
\end{align}

\medskip
In order to find $R$ and $T$, we used the definition of the electrical current density for the one-dimensional Klein-Gordon equation \ref{klein}

\begin{equation}
\label{current}
\vec{j}=\frac{i}{2}\left(\phi^*\vec{\nabla}\phi-\phi\vec{\nabla}\phi^*\right). \tag*{(23)}
\end{equation}

\medskip
The current as $x \rightarrow -\infty$ can be decomposed as $j_\textnormal{L}=j_\textnormal{inc}-j_\textnormal{refl}$ where $j_\textnormal{inc}$ is the incident current and $j_\textnormal{ref}$ is the reflected one. Analogously we have that, on ther right side, as $x \rightarrow \infty$ the current is $j_\textnormal{R}=j_\textnormal{trans}$, where $j_\textnormal{trans}$ is the transmitted current \cite{rojas:2005}.

\medskip
The reflection coefficient $R$, and the transmission coefficient $T$, in terms of the incident $j_\textnormal{inc}$, reflected $j_\textnormal{ref}$, and transmission $j_\textnormal{trans}$ currents are

\begin{equation}
\label{R}
R=\frac{j_\textnormal{ref}}{j_\textnormal{inc}}=\frac{|B|^2}{|A|^2}. \tag*{(24)}
\end{equation}

\begin{equation}
\label{T}
T=\frac{j_\textnormal{trans}}{j_\textnormal{inc}}=\frac{\mu}{\nu}\frac{1}{|A|^2}. \tag*{(25)}
\end{equation}

\medskip
The reflection coefficient $R$, and the transmission coefficient $T$ satisfy the unitary relation $T+R=1$ and are expresses in terms of the coefficients $A$ and $B$, therefore are expressed in terms of the Gamma function and they are determined with te software Maple 18. 

The dispersion relation $\nu$ and $\mu$ must be positive because it correspond to an incident particle moved from left to right and,  their sign depends on the group velocity, define by \cite{calogeracos:1999}

\begin{equation}
 \label{group_nu}
 \frac{dE}{d\nu'}=\frac{\nu'}{E+a}\geq 0. \tag*{(26)}
\end{equation}

\begin{equation}
 \label{group_muu}
 \frac{dE}{d\mu'}=\frac{\mu'}{E-a}\geq 0. \tag*{(27)}
\end{equation}

\medskip
For the hyperbolic tangent potential we have five different regions, this regions are observed in table \ref{regions}.

\bigskip
\begin{table}[h]
\begin{center}
\begin{tabular}{c|c|c|c|c}
\hline\hline
$E>a+m$       & $\nu'>0$  &  $\nu\in\Re$  & $\mu'>0$  &  $\mu\in\Re$\\
$a+m>E>a-m$   & $\nu'>0$  &  $\nu\in\Re$  &           &  $\mu\in\Im$\\
$a-m>E>-a+m$  & $\nu'>0$  &  $\nu\in\Re$  & $\mu'<0$  &  $\mu\in\Re$\\
$-a+m>E>-a-m$ &           &  $\nu\in\Im$  & $\mu'<0$  &  $\mu\in\Re$\\
$E<-a-m$      & $\nu'<0$  &  $\nu\in\Re$  & $\mu'<0$  &  $\mu\in\Re$\\
\hline\hline
\end{tabular}
\caption{\label{regions}Regions for $\nu'$ and $\mu'$.}
\end{center}
\end{table}

\medskip
It is important to note that in the regions $a+m>E>a-m$ and $-a+m>E>-a-m$ the dispersion relations $\mu$ and $\nu$ are imaginary pure and the transmitted wave if attenuates, so $R=1$. In the region $a-m>E>-a+m$, $\mu'<0$ and, $\nu'>0$ we have that $R>1$, so superradiance occurs.

\begin{figure}[!th]
\begin{center}
\subfigure[]{\includegraphics[scale=0.30]{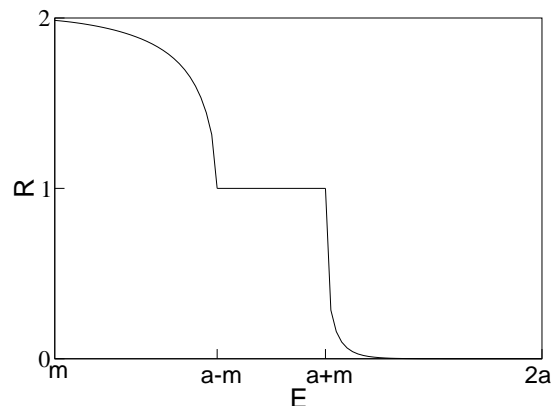}}
\hspace{1cm}\subfigure[]{\includegraphics[scale=0.30]{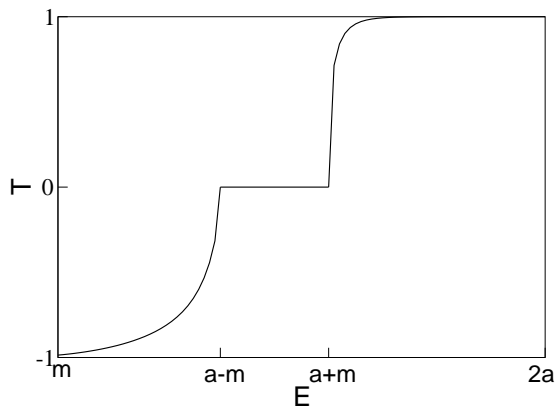}}
\caption{\label{TandR_b=2} The reflection $R$ and transmission $T$ coefficients varying energy $E$ for the relativistic hyperbolic tangent potential for $a=5$, $b=2$ and, $m=1$.}
\end{center}
\end{figure}

\begin{figure}[!th]
\begin{center}
\bigskip
\subfigure[]{\includegraphics[scale=0.30]{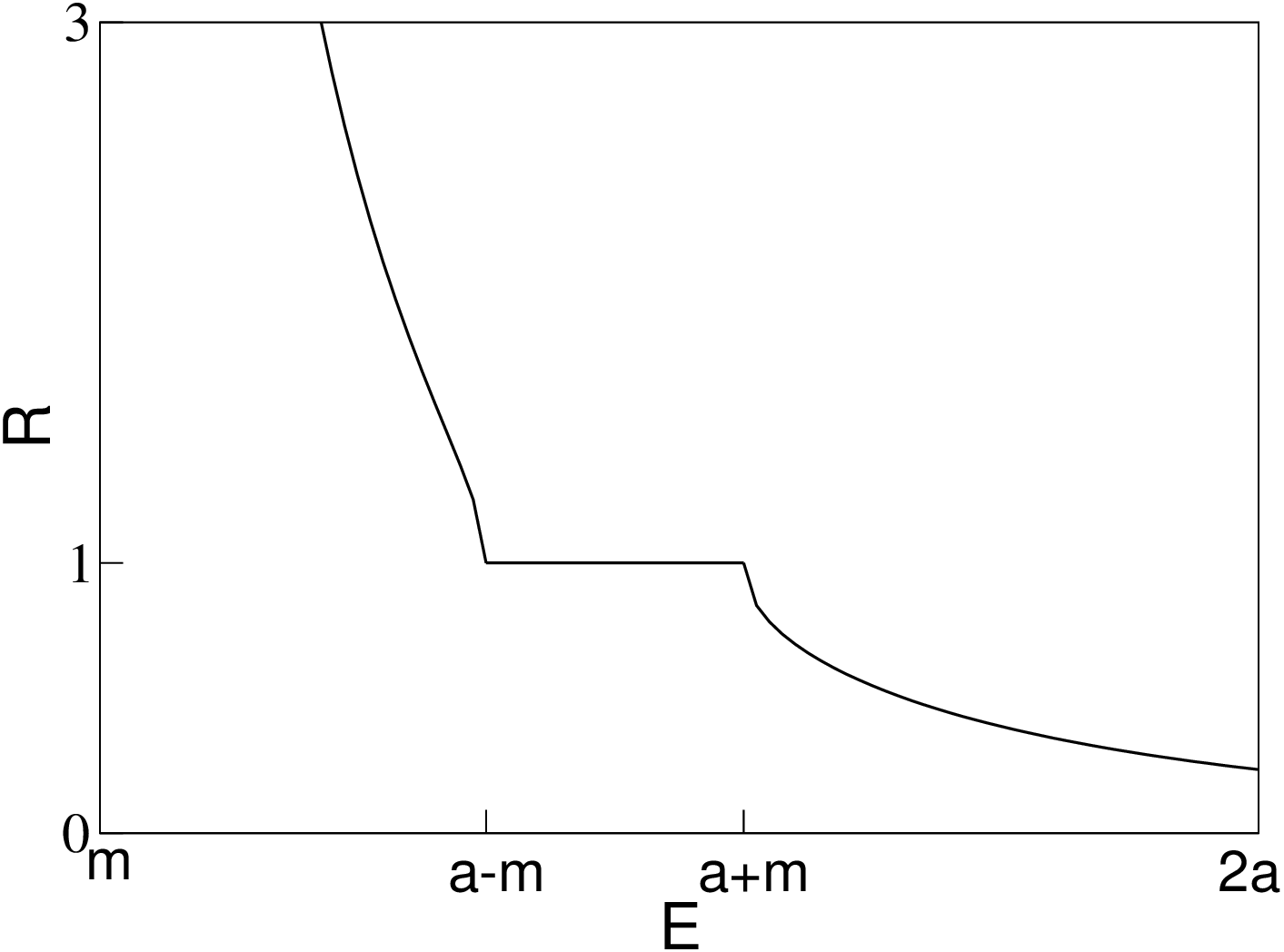}}
\hspace{1cm}\subfigure[]{\includegraphics[scale=0.30]{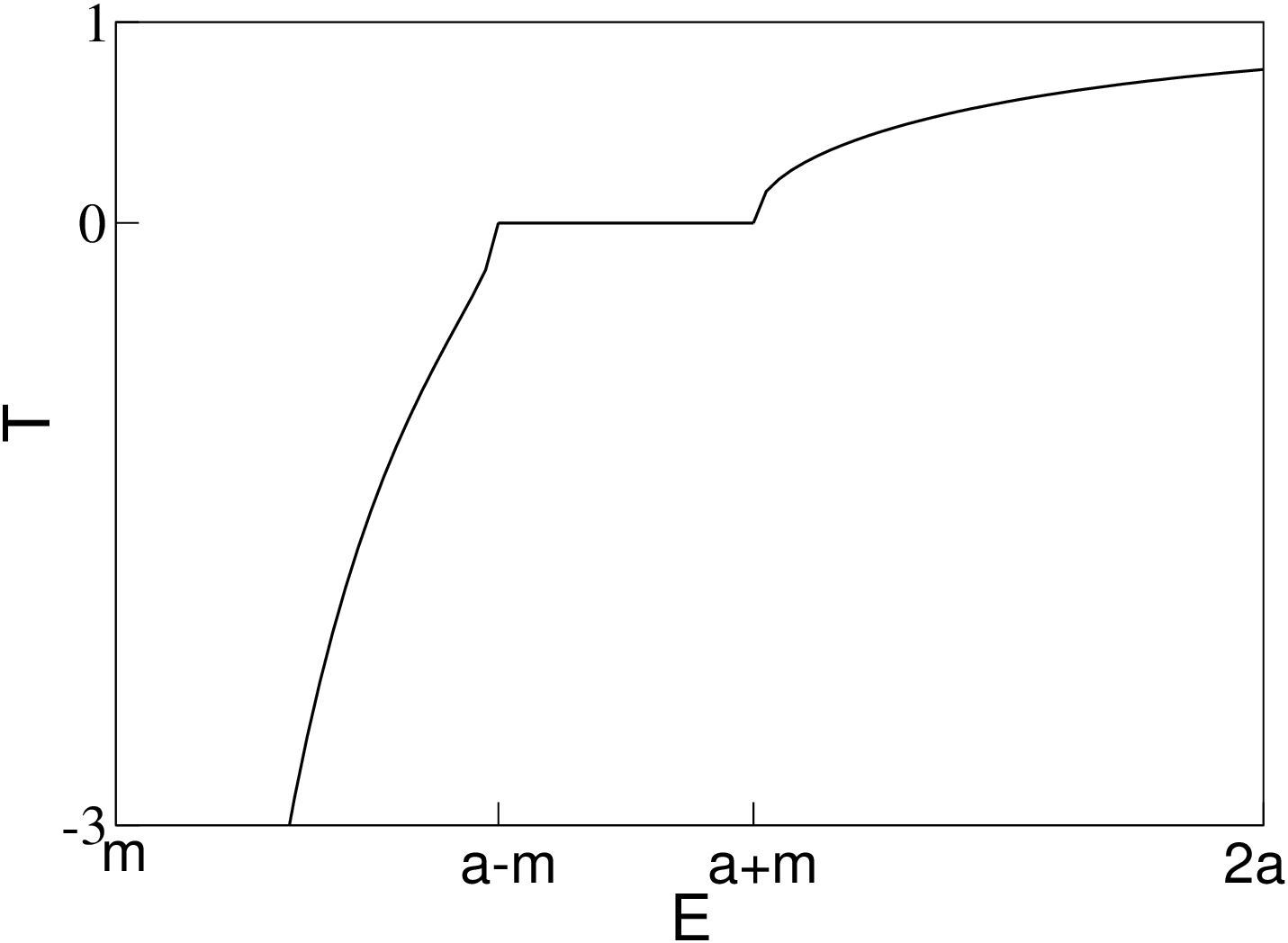}}
\caption{\label{TandR_b=50} The reflection $R$ and transmission $T$ coefficients varying energy $E$ for the relativistic hyperbolic tangent potential for $a=5$, $b=50$ and, $m=1$.}
\end{center}
\end{figure}

\medskip
Figs. \ref{TandR_b=2}(a) and \ref{TandR_b=2}(b)  show the  reflection $R$ and transmission $T$ coefficients for $E>m$, $a=5$ and $b=2$. Figs. \ref{TandR_b=50}(a) and \ref{TandR_b=50}(b) show the  reflection $R$ and transmission $T$ coefficients $R$ for $E> m$, $a=5$ and $b=50$. 
We observed in the figures that in the region $a-m>E>m$ the reflection coefficient $R$ is bigger than one whereas the coefficient of transmission $T$ becomes negative, so we observed superradiance \cite{cheng:2009, wagner:2010} and that the coefficients $R$ and $T$  satisfy the unitary condition  $T+R=1$. The hyperbolic tangent potential is useful to study the superradiance phenomenon.

\section{Conclusion}
\label{conclusion}
In this paper we have discussed the scattering solutions of the Klein-Gordon equation in presence of the hyperbolic tangent potential. The solutions are determined in terms of hypergeometric function. The reflection $R$ and transmission $T$ coefficients are determined in terms of the Gamma function. We have shown that for the region where $a-m>E>m$, the phenomenon of superradiance occurs.


\end{document}